\documentclass[11pt]{article}
\usepackage{geometry}
\usepackage{svg}
\usepackage[tight]{subfigure}
\usepackage{float}
\usepackage[affil-sl ]{authblk}
\usepackage{hyperref}

\let\OLDthebibliography\thebibliography
\renewcommand\thebibliography[1]{
	\OLDthebibliography{#1}
	\setlength{\parskip}{0pt}
	\setlength{\itemsep}{0pt plus 0.3ex}
}

\usepackage{fancyhdr}
\fancypagestyle{spin}{%
	\fancyhf{}%
	\fancyhead[]{}%
	\fancyfoot[R]{\thepage}
	\fancyfoot[L]{SPIN2021 - 24th International Spin Symposium}
}
\title{Polarized Electron Beams from Laser Plasma Acceleration and Their Polarimetry}

\author[1,2]{Jennifer Popp\footnote{E-mail: jennifer.popp@desy.de}}
\author[1]{Simon Bohlen}
\author[1,2]{Felix Stehr}
\author[1]{Jenny List}
\author[2,1]{Gudrid Moortgat-Pick}
\author[1]{Jens Osterhoff}
\author[1]{Kristjan P\~oder}
\affil[1]{Deutsches Elektronen-Synchrotron DESY, Notkestr. 85, 22607 Hamburg, Germany}
\affil[2]{Universität Hamburg, Department of Physics, Jungiusstr. 9, 20355 Hamburg, Germany}

\date{February 10, 2022}

\begin{document}
	\maketitle

	\begin{abstract}
		In recent years, Laser Plasma Acceleration (LPA) has become a promising alternative to conventional RF accelerators. However, so far, it has only been theoretically shown that generating polarized LPA beams is possible. 
		The LEAP (Laser Electron Acceleration with Polarization) project at DESY aims to demonstrate this experimentally for the first time, using a pre-polarized plasma target.\\
		The electron polarization will be measured with photon transmission polarimetry, which makes use of the production of circularly polarized bremsstrahlung during the passage of the electron beams through a suitable converter target. The photon polarization is then measured with the aid of transmission asymmetry arising from reversing the magnetization direction of an iron absorber.\\
		In this contribution an overview of the LEAP project is presented, detailing the generation of the polarized electron beams along with the design and simulation studies of the polarimeter.
	\end{abstract}

	\section{Introduction}
	Polarized particle beams are a key instrument for the investigation of
	spin-dependent processes and are therefore indispensable for many experiments in particle, atomic and nuclear physics \cite{Buescher20}. 
	Conventional sources, like the photoemission from GaAs as planned eg. at the ILC \cite{ILCTDR}, realise bunches containing $3 \times 10^{10}$ electrons and a polarization of 90\,\%, but then rely on RF cavities, limited by material breakdown thresholds, for further acceleration.\\
	In contrast, plasma accelerators, which are not impaired by this limit, can sustain acceleration gradients more than three orders of magnitude higher \cite{Esaray09} and therefore, offer a compact alternative.
	Furthermore, the short bunch length of the electron beams generated by the plasma accelerators of only a few femtoseconds can be advantageous for imaging applications and they have an inherent synchronization to the laser driver, which is needed for pump and probe experiments.\\
	So far the influence of plasma acceleration on the electron polarization has only been studied theoretically. These studies suggest that the polarization can be maintained \cite{Viera11} and possibilities of generating polarized particle beams with Laser Plasma Acceleration (LPA) have been investigated using simulations \cite{Wen19,Wu19}.
	LEAP is a proof of principle experiment at DESY aiming to demonstrate the latter. 
	This paper gives an overview of the LEAP project, detailing the generation of polarized beams from LPA \cite{Tajima1979}, simulation studies, and a design for the polarimeter. 
	
	\section{Laser Plasma Acceleration of Polarized Electron Beams}
	In an LPA, a relativistically intense laser pulse (in the order of $10^{18}$\,W/cm$^2$ for a Ti:Sapphire laser system) is incident on a plasma source. The ponderomotive force, proportional to the intensity gradient of the laser pulse, acts on all charged particles in a plasma with a strength that is proportional to the particle's charge and inversely proportional to their mass.  This means that electrons are pushed out of the highest intensity regions of the laser, but ions, due to their higher mass, remain initially immobile. The displacement of the electrons creates a restoring force and the electrons start to oscillate at the characteristic electron-plasma frequency and form a wave like structure. The resulting longitudinal fields in the plasma wave can reach more than 100\,GV/m. Electrons injected at the back of this wave can be accelerated in the wake field of the laser pulse. 
	
	\subsection{Polarized beams from a pre-polarized gas jet target}
\begin{figure}[H]
		\centering
		\subfigure[][]{
			\label{leap_proc}
			\centering
			\includegraphics[width=0.8\textwidth]{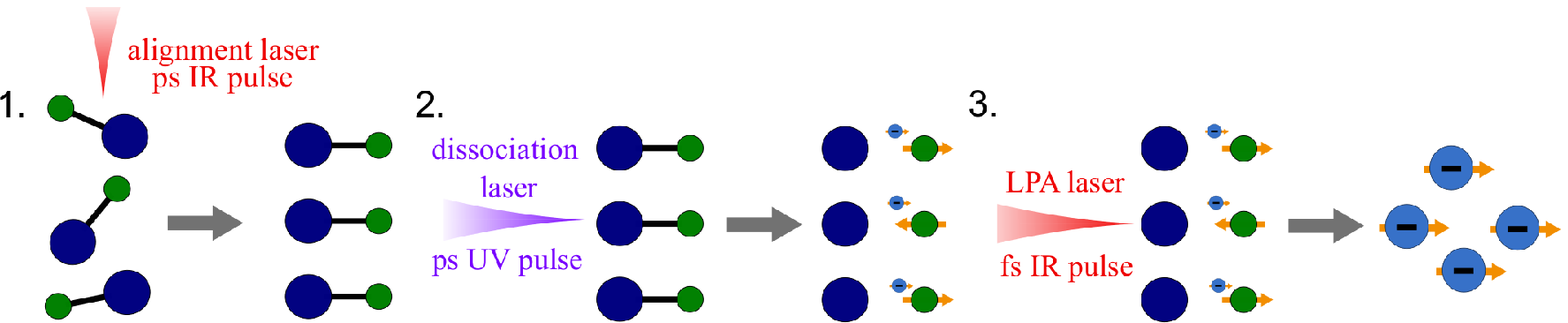}
		}\\
		\vspace{-11 pt}
		\subfigure[][]{
			\label{acc_setup}
			\centering
			\includegraphics[width=0.8\textwidth]{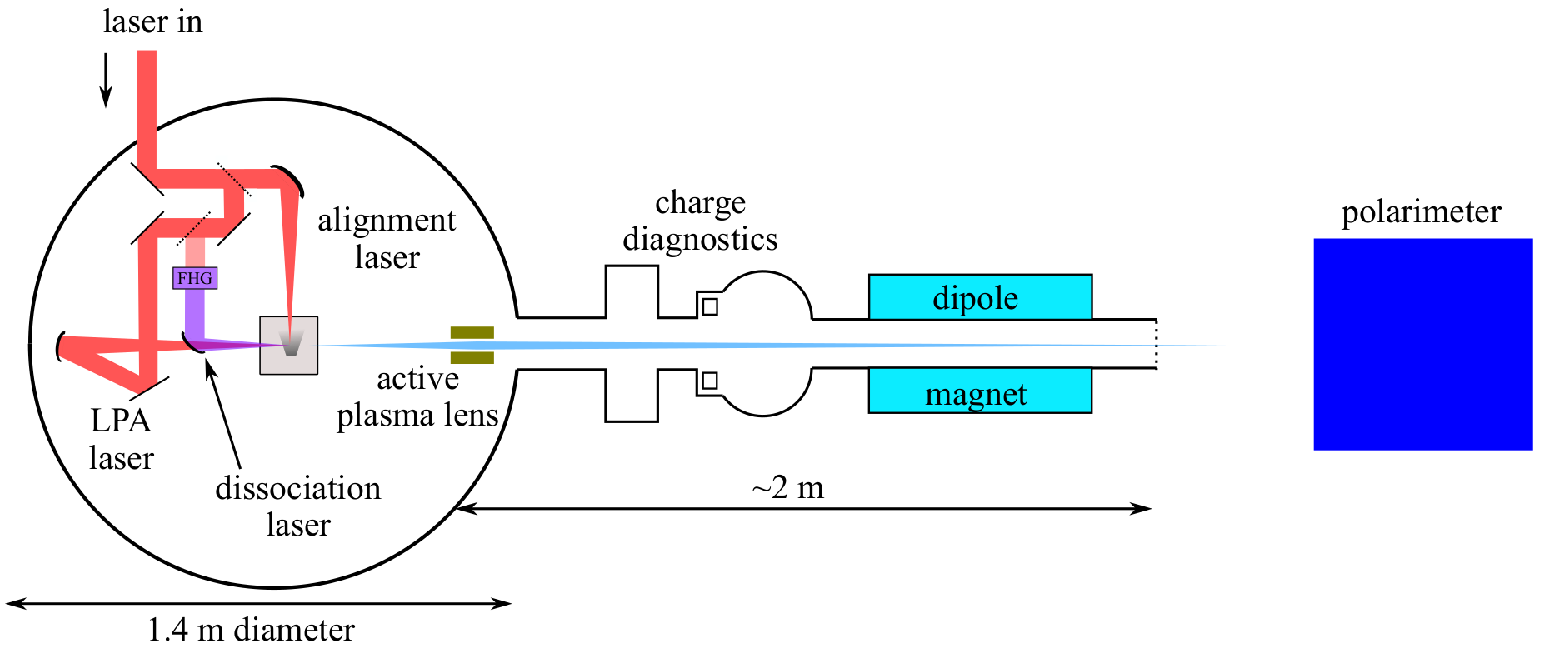}		
		}
		
		\caption{a) Schematic depiction of the pre-polarization and acceleration in a polarized LPA. b) Schematic layout of the LEAP experimental setup.}
	\end{figure}

	One proposed method to generate polarized beams with LPA is to use a pre-polarized gas jet plasma source \cite{Spiliotis21,Wen19}. This method is depicted in Fig.\,\ref{leap_proc} and works as follows: 
	First, the molecular bond of diatomic gas molecules is aligned in the electric field of a laser pulse. This is followed by the photodissociation into molecular states with polarized valence electrons using a circularly polarized UV laser pulse of certain wavelength \cite{Spiliotis21}. The third pulse then drives a plasma wave and injected electrons are accelerated. Suitable regimes to maintain the beam polarization during the acceleration have been found in simulations \cite{Viera11,Fan22}.       
	
	\subsection{Setup at DESY}
	A robust and stable LPA source employing a 10 Hz, 25 TW Ti:Sapphire laser system has already been commissioned and has been used as the electron source for the novel PLASMED\,X all-optical inverse Compton source for medical applications. Stable acceleration of electron bunches over several hours was demonstrated \cite{Bohlen21}. For the LEAP project, the existing setup will be modified to enable the demonstration of polarized electron beams from a LPA.\\
	The laser pulse is split into three parts as indicated in Fig \ref{acc_setup}. The first part is stretched to picosecond duration and used for aligning the molecules. The second part undergoes a cascaded second harmonic generation to generate a fourth harmonic UV pulse needed for the dissociation process.  The remainder of the laser is used for injection and acceleration of the polarized electrons. An active plasma lens is planned to be used for beam capture.\\ 
	Based on previous measurements \cite{Bohlen21} and particle-in-cell simulations using FBPIC \cite{Lehe16}, the following beam parameters are expected: A tunable central energy from 30 to 70\,MeV, a FWHM energy spread of about 10\,\%, an electron spot size of 1\,{\textmu}m, a divergence of 3\,mrad, and an electron beam charge of 10\,pC.\\ 
	The choice of target gas is an important consideration for any polarized LPA source employing a pre-polarized plasma source as it defines the achievable fraction of polarized electrons and the wavelength required for the dissociation of the molecules. As only the electrons of the dissociated molecular bond can be pre-polarized, the maximum fraction of theoretically polarized electrons $P_{\mathrm{max}}$ is limited.\\
	Hydrogen chloride has many electron levels and, depending on how many are ionized, a maximum polarization between 12.5\% and 25\,\% is feasible.
	Hydrogen fluoride, which offers the second highest maximum polarization of 25\,\% requires a dissociation laser with wavelength of less than 150 nm which is difficult to achieve and forms hydrofluoric acid when coming into contact with moisture and is therefore no option.
	Other hydrogen halides would lead to a lower $P_{\mathrm{max}}$ than that of HCl.
	Hydrogen, has a maximum plasma source pre-polarisation of 100\,\%. However, choosing pure $\mathrm{H}_2$ as target gas, has the drawback that the wavelength required for the dissociation of around 100 nm is challenging to generate and currently not yet practicable.
	
	\section{Polarimetry of laser-plasma-accelerated electron beams}
	In principle any well-known, high rate and polarization dependent process can be used for polarimetry. Typical choices comprise Mott or M{\o}ller scattering on thin targets or laser Compton scattering. However, none of these work well at the energy range of LEAP:
	For Mott scattering the cross-section and the scattering angles become impractically small above ~10\,MeV. M{\o}ller scattering starts to become practical for energies above ~200\,MeV and Compton scattering requires at least a few GeV, as otherwise the asymmetry becomes inconveniently small to detect \cite{Schaelicke07a}.   
	Thus photon transmission polarimetry, which relies on Compton scattering of  photons on electrons in a magnetized material, is the method that is best suited for the expected energy range of several tens of MeV. This method has been successfully used for the polarimetry of 4-8 MeV positrons at the E166 experiment \cite{Alexander09}, where a polarization of about 80\,\% was measured with a relative uncertainty of about 10-15\,\%. 
	
	\subsection{Transmission Polarimetry}
	Photon transmission polarimetry, as depicted in Fig.\,\ref{pol_setup} is divided into three steps: Firstly, the conversion of electrons to photons, secondly, the transmission depending on the polarization and finally, the photon detection.
	When the electron bunch polarized in its propagation direction is irradiated onto a suitable converter target (e.g. several millimeter of tungsten) circularly polarized photons are created via bremsstrahlung \cite{Likhachev02}. A typical spectrum peaks in the range of keV and has a vanishing tail at the beam energy.\\
	These photons then traverse through a magnetized block of iron. 
	The total material interaction coefficient decreases with higher photon energy, i.e. a high energy photon is more likely to traverse the material without interacting. The polarization dependent part of the number of transmitted, or rather Compton scattered in forward direction, photons is proportional to $\exp{(-P_{\mathrm{\gamma}}P_{\mathrm{e^-}})}$, where $P_{\mathrm{e^-}}$ is the polarization of the free electrons in the iron block and $P_{\mathrm{\gamma}}$ the photon polarization. 
	The number of photons exiting the iron block is largest when $P_{\mathrm{e^-}}$ and $P_{\mathrm{\gamma}}$ are antiparallel.\\
	Photon detection will utilize a calorimeter consisting of nine lead glass crystals, attached to photomultiplier tubes PMTs, wrapped in aluminum foil and arranged in a $3\times3$ grid. From the measured PMT signal the sum of the energy of the photons transmitted through the magnet can be inferred.\\
	When the direction of the magnetic field is changed, $P_{\mathrm{e^-}}$ changes accordingly, leading to a difference in photon transmission rate and hence, in measured photon energy sum. An asymmetry can be defined as the difference between the transmitted photon energy for anti-parallel and parallel polarization directions divided by the sum of the two:
	\begin{equation}
		\delta = \frac{E_{AP} - E_{P} }{E_{AP} + E_{P} }
	\end{equation}
	The magnitude of the asymmetry with respect to the magnetization direction is proportional to the photon polarization.
	
	\subsection{Simulations for setup optimization}
\begin{figure}[H]
		\begin{centering}
			\subfigure[][]{
				\centering
				\includegraphics[width=0.7\textwidth]{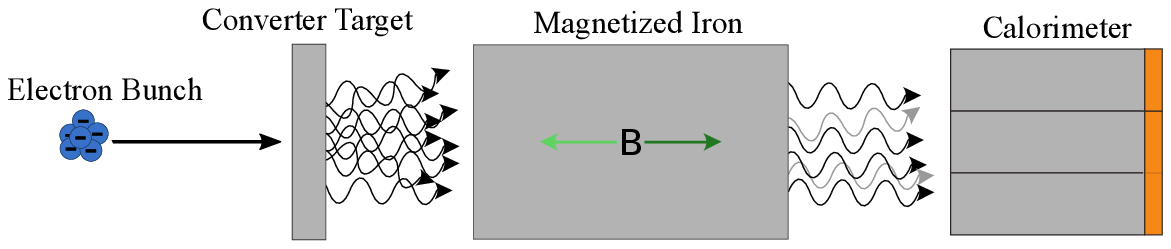}
				\label{pol_setup}
			} \\ 
		\end{centering}
		\vspace{-11 pt}
		\subfigure[][]{
			\centering
			\includegraphics[width=0.49\textwidth]{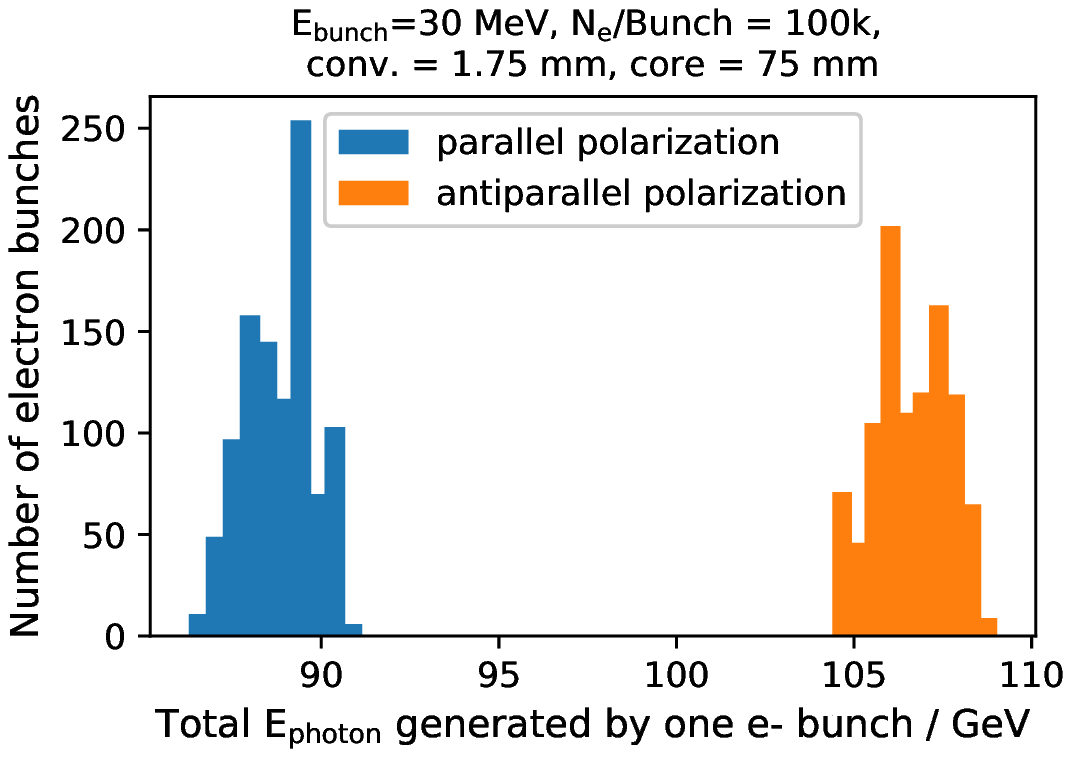}
			\label{esums}
		}
		\hspace{8pt}
		\subfigure[][]{
			\centering
			\includegraphics[width=0.49\textwidth]{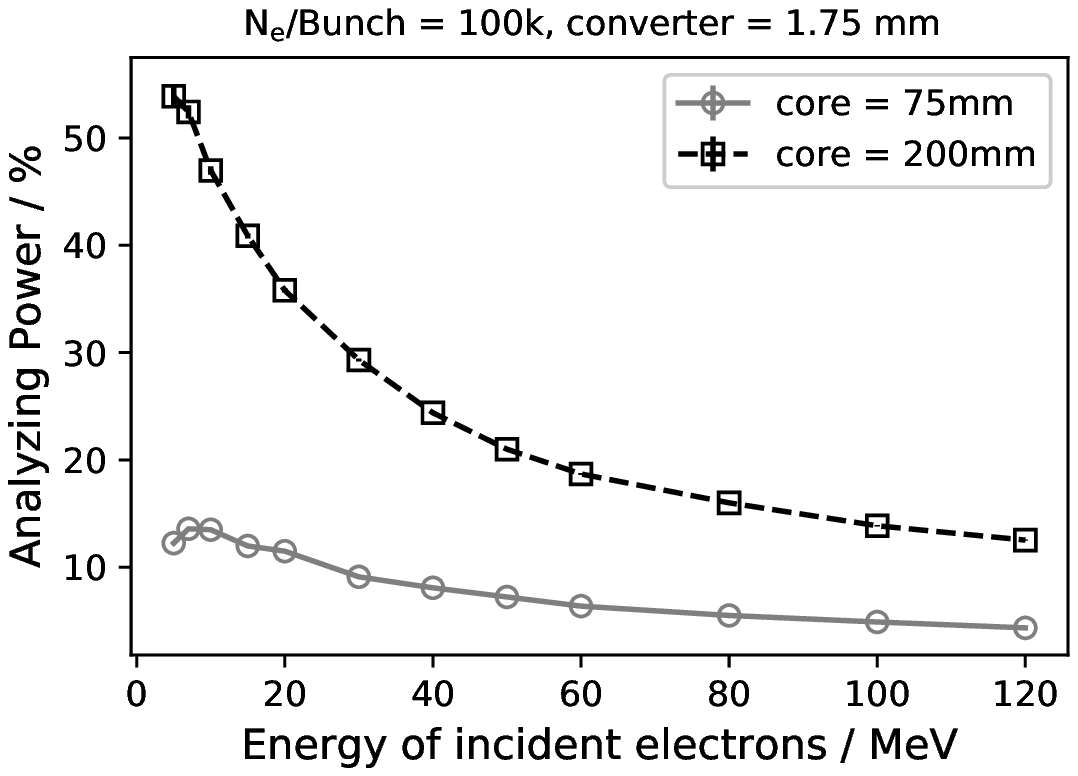}
			\label{eruns}
		}
		\caption{a) Working principle of photon transmission polarimetry: Production of polarized Bremsstrahlung followed by polarization direction depended transmission through a magnetized absorber. b) Distribution of total photon energies for 100k electron bunches c) Analyzing power for various electron bunch energies.}
		\label{f2}
	\end{figure}

	To optimize the polarimeter for the LEAP setup, the polarimeter setup has been implemented in a Geant4 simulation \cite{Agostinelli03}.
	As a first step, mono-energetic, simplified particle beams were simulated.  Geant4 physicslists were used that include polarized Compton scattering, $\gamma$-conversion, ionization, bremsstrahlung, positron annihilation and photoelectric effect \cite{Schaelicke07b}.
	The analyzing power $A_{\mathrm{e}}$ was obtained as the expected asymmetry for 100\% polarized beam and absorber by setting the polarizations of the electron beam and iron core to $\pm 1$. The experimentally measured asymmetry can be divided by $A_{\mathrm{e}}$ and the absorber magnetization to yield the actual electron beam polarization.\\ 
	The analyzing power is calculated with the aid of the average total photon energy per electron bunch. An example of the distribution of those energies is shown in Fig. \ref{esums}. The energy dependence of the analyzing power is shown for two different core lengths in Fig. \ref{eruns}. It is evident that for a fixed absorber length, $A_{\mathrm{e}}$ decreases with increasing electron energy. Therefore, lower electron energies are more promising for transmission polarimetry and therefore more suited for demonstration experiments such as LEAP. From Fig. \ref{eruns} it can also be seen, that $A_{\mathrm{e}}$ increases with core thickness, eg. from 10\,\% at 25 MeV and 75 mm to 32\,\% at a core thickness of 200 mm. The iron thickness needs to be matched to the energy of the incident photons.
	\section{Summary}
	Polarized particle beams are an important tool to investigate spin dependent processes and plasma accelerators are a promising candidate for compact machines in the future. It has yet to be shown that polarized beams can be produced and accelerated in plasma accelerators.
	LEAP is a proof-of-principle experiment under construction at DESY aiming to demonstrate LPA with a prepolarized plasma source utilizing photon transmission polarimetry. Design studies for setup optimization and calorimeter tests are ongoing. 
	\section*{Acknowledgements}
	This work has benefited from computing services provided by the German National Analysis Facility (NAF).
	The authors gratefully acknowledge funding from the DESY Strategy Fund and the Gauss Centre for Supercomputing e.V. for funding this project by providing computing time on the GCS Supercomputer JUWELS at Jülich Supercomputing Centre (JSC).
	The authors also would like to thank Karim Laihem for providing code.


\begin{thebibliography}{9}
		\bibitem{Buescher20} M. B\"{u}scher, A. H\"{u}tzen, L. Ji and A. Lehrach, High Power Laser Sci. Eng. \textbf{8}, E36 (2020)
		\bibitem{ILCTDR} ILC TDR Volume 3.II, Accelerator Baseline Design (2013).
		\bibitem{Esaray09} E. Esaray, C. Schroeder and W. Leemans, Rev. Mod. Phys. \textbf{81}(3), 1229 (2009).
		\bibitem{Viera11}J. Viera et al., Phys Rev. ST Acell. Beams \textbf{14}, 071303 (2011)
		\bibitem{Wen19}M. Wen, M. Tamburini and H. Keitel, Phys. Rev. Lett. \textbf{122},214801 (2019)
		\bibitem{Wu19}Y. Wu et al.,Phys Rev. E \textbf{100}, 043202 (2019)
		\bibitem{Tajima1979} T. Tajima and J. M. Dawson, Phys. Rev. Lett. \textbf{43}, 267 (1979)
		\bibitem{Fan22}	H. C. Fan et al., arXiv:2201.02969 \textbf{[physics.plasm-ph]}(2022)
		\bibitem{Bohlen21}S. Bohlen et al., arXiv:2203.00561 \textbf{[physics.acc-ph]} (2022)
		\bibitem{Lehe16} R. Lehe et al. (2016). CPC 203, P. 66-82   
		\bibitem{Spiliotis21}A. Spiliotis et al., Light Sci. Appl. \textbf{43}(35), (2021)
		\bibitem{Schaelicke07a}A. Sch\"{a}licke et al., Pramana - J. Phys. \textbf{69}, 1171-1175 (2007)
		\bibitem{Alexander09}G. Alexander et al., Nucl. Instrum. Methods Phys. Res. A, \textbf{610}, 2, 451-487 (2009)
		\bibitem{Likhachev02}Likhachev et al., Nucl. Instrum. Methods Phys. Res. A, \textbf{495}, 2, 139-17 (2002)
		\bibitem{Agostinelli03}S. Agostinelli et al., Nucl. Instrum. Meth. A \textbf{506},250-303 (2003) 
		\bibitem{Schaelicke07b}A. Sch\"{a}licke, K. Laihem, P. Starovoltov, arXiv:0712.2336 (2007)
		
	\end{thebibliography}
\end{document}